\documentstyle[12pt,subeqn,subeqnarray]{article}

\catcode `@=11 \@addtoreset{equation}{section} \catcode `@=12
\begin{document}
\def\be{\begin{equation}}
\def\ee{\end{equation}}
\def\ba{\begin{array}{l}}
\def\ea{\end{array}}
\def\beq{\begin{eqnarray}}
\def\eeq{\end{eqnarray}}
\def\eq#1{(\ref{#1})}
\def\del{\partial}
\def\A{{\cal A}}

\renewcommand\arraystretch{1.5}

\begin{flushright}
TIFR-TH-98/05\\
February 1998
\end{flushright}
\begin{center}
\vspace{3 ex}
{\large{\bf PROBING 4-DIMENSIONAL NONSUPERSYMMETRIC BLACK HOLES
CARRYING D0- AND D6-BRANE CHARGES}} \\
\vspace{8 ex}
Avinash Dhar$^*$ and Gautam Mandal$^\dagger$ \\
{\sl Department of Theoretical Physics} \\
{\sl Tata Institute of Fundamental Research,} \\
{\sl Homi Bhabha Road, Mumbai 400 005, INDIA.} \\
\vspace{15 ex}
\pretolerance=1000000
\bf ABSTRACT\\
\end{center}
\vspace{2 ex}

We discuss a 4-dimensional nonsupersymmetric black hole solution to
low energy type IIA string theory which carries D0- and D6-brane
charges. For equal charges this solution reduces to the one discussed
recently by Sheinblatt. We present a new parametrization of the
solution in terms of four numbers which reveals the underlying brane
and antibrane structure of the black hole arbitrarily far from
extremality. In this parametrization, the entropy of the 
general nonextremal black hole takes on a simple U-duality
invariant form. A Yang-Mills solution for the brane configuration
corresponding to the extremal case is constructed and a computation of
the 1-loop matrix theory potential for the scattering of a 0-brane
probe off this brane configuration done. We find that this agrees with
the 1-loop potential obtained from a supergravity calculation in the
limit in which the ratio of the 0-brane to 6-brane charges is large.

\vfill
\hrule
\vspace{0.5 ex}
\leftline{$^*$ adhar@theory.tifr.res.in}
\leftline{$^\dagger$ mandal@theory.tifr.res.in}
\clearpage

\section{INTRODUCTION}

In recent years spectacular progress has been made towards a
microscopic derivation of black hole thermodynamics based on string
theory models \cite{one, two, three, four, five, six, seven,
eight}. \footnote{For a recent review and a more complete list of
references see \cite{eighta}.}
Central to this development has been 
the existence of Dirichlet p-brane solitons (Dp-branes) of string
theory, and a description of their dynamics in terms of open strings.
For recent reviews of this area we refer the reader to \cite{nine, ninea}.
One of the remarkable and unexpected consequences of this activity has
been the uncovering of a deep and potentially far-reaching
connection between supergravity and superYang-Mills theories
\cite{twelve, ten, eleven}. A precise formulation of this connection
exists in the form of the matrix theory conjecture \cite{thirteen}.
Although there are several issues connected with this conjecture that
need better understanding, there is an impressive body of evidence in
support of it.  For recent reviews in this area we refer the reader to
\cite{fourteen, ninea}.

In the context of black hole physics, there have been many studies
which explore this supergravity--superYang-Mills connection. Many of
these studies use D-branes to probe black hole physics \cite{fifteen,
sixteen, seventeen, eighteen, nineteen} 
because slowly moving D-branes act as localized probes and so can be
used for big black holes, where weak coupling perturbation theory
around the classical supergravity solution is good, as well as for
small size ($<$ string length, $l_s$) black holes, where perturbative
D-brane gauge theory is good. In all the known cases agreement is
found at 1-loop level between the supergravity and superYang-Mills
calculations. 

All the examples of black holes quoted above are either extremal BPS
or near extremal BPS.\footnote{Similar agreement at 1-loop level has also
been found for D-brane scattering off other supersymmetric or nearly
supersymmetric bound states of D-branes \cite{twentyone}.} In these cases
there is either some residual supersymmetry or there is a small
parameter which controls deviations from a configuration that
preserves some supersymmetry. One might argue that it is this fact
that is responsible for the agreement mentioned above as well as the
agreement of other physical quantities like entropy, etc.

It is clearly of interest to ask what happens in the case of
nonsupersymmetric black holes, i.e. those that neither preserve any
supersymmetry nor are in any obvious sense close to one that does. The
matrix theory conjecture requires agreement between supergravity and
superYang-Mills calculations even in these cases.\footnote{Agreement has
been shown for the nonsupersymmetric configuration of a membrane and
anti-membrane \cite{twentytwo}. The authors of this work have argued that in
the limit of large boosts, required by the matrix theory conjecture,
this nonsupersymmetric configuration comes close to being a
supersymmetric one.} \footnote{Other examples of agreement exist. For a
certain class of black holes far from extremality microscopic counting
agrees with the Bekenstein-Hawking formula \cite{twentythree}. Microscopic
counting for a nonsupersymmetric extremal black hole in Type IA theory
also reproduces the Bekenstein-Hawking entropy exactly \cite{twentyfour}.
Similar agreement has been seen for entropy at the stretched horizon
in the case of some nonsupersymmetric extremal elementary black holes
\cite{twentyfive}. Also, recently there has been a lot of activity in
identifying Schwarzschild black holes in the matrix theory
\cite{twentysix, twentyseven, twentyeight, twentynine, thirty, thirtyone}.}
It is important to verify if the conjecture is right in these cases
also, since this would provide additional information about strongly
coupled dynamics that does not entirely rely on supersymmetry. 
One such nonsupersymmetric extremal black hole solution to the 
classical low energy equations of Type IIA theory compactified down to
4-dimensions has recently been discussed by several authors
\cite{thirtytwo, thirtythree, thirtyfour, thirtyfive}. This
4-dimensional black hole carries D0- and D6-brane charges 
and a qualitative microscopic picture of it as a bound state 
of D0- and D6-branes
has been developed in \cite{thirtytwo}. The solution given in this work is,
however, restricted to the case in which the D0-brane charge is equal
to the D6-brane charge. In this solution the D0-brane charge cannot be
varied independently and so one cannot go to the infinite momentum
frame, as required by the matrix theory conjecture. Thus, for comparison 
with matrix theory we need to generalize the
above known solution to the one in which the D0- and D6-brane charges
can be varied independently.\footnote{See, however,
reference \cite{thirtyfour} in which a matrix theory calculation in 
the related problem of scattering
of probe D6-branes carrying large D0-brane charge off a target
D0-brane shows agreement with supergravity calculation at 1-loop.}

In this paper we will discuss this generalization and compare the
supergravity and superYang-Mills calculations of the effective
potential at 1-loop for a D0-brane probe scattering off the black
hole. The black hole solution is discussed in Sec. 2. A paramerization
of the solution in terms of four numbers is presented in Sec. 3. In
terms of these numbers the entropy of the black hole takes on a very
simple form, even far away from extremality, revealing an underlying
brane and antibrane structure.  In Sec. 4 we study a slowly moving
D0-brane probe in the presence of the extremal black hole background
from the low energy classical closed string point of view. In Sec. 5
we generalize the Yang-Mills construction of bound state of D0- and
D6-branes of Taylor \cite{thirtythree} to the present solution. The
1-loop matrix theory calculation is done in Sec. 6 and compared with
the supergravity calculation of Sec. 4. We conclude with some remarks
in Sec. 7.

After the first version of this work was submitted we learnt of a
related work \cite{thirtyfivea} with which we have some overlap.

\section{THE BLACK HOLE SOLUTION}

The bosonic part of the low energy effective action of Type IIA string
theory is
\be
S_{10} = \frac{1}{(2\pi)^7 g^2} \int d^{10} x \sqrt{-G_{10}} \left[
e^{-2\phi} (R_{10} + 4 (\nabla \phi)^2) - \frac{1}{4} F^2_{10} \right]
\ee
We have set all the matter terms, except the 2-form Ramond-Ramond
field strength $F_{10}$, to zero. The string coupling $g$ is defined
such that the dilaton field $\phi \rightarrow 0$ at spatial infinity.
Also, we have used the signature $(-, +, + , \cdots )$ and string
units $\sqrt{\alpha'} = l_s = 1$.  The 10-dimensional Newton's constant
is then given by $8\pi^6g^2$ so that the overall constant in front in
the action (2.1) is $1/16 \pi $(Newton's constant). As discussed in
\cite{thirtysix}, a solution to the classical equations of motion of the
action (2.1) can be obtained from a solution to the classical
equations of motion of the bosonic part of 11-dimensional supergravity action
\be
S_{11} = \frac{1}{(2\pi)^8 g^3} \int d^{11} x \sqrt{-G_{11}} R_{11}
\ee
of the form
\be
ds^2_{11} = e^{4\phi/3} \left(dx_{11} + A_\mu dx^\mu \right)^2 +
e^{-2\phi/3} ds^2_{10} ,
\ee
by compactifying $x_{11}$ on a circle of radius $g$, provided
$\partial/\partial x_{11}$ is a Killing vector of the solution. The
solution for the various fields can then be read-off from the from of
the 11-dimensional line element in (2.3). Here $A_\mu$ is
10-dimensional Ramond-Ramond gauge potential from which the field
strength $F_{10}$ is derived.

We are interested in a 4-dimensional solution of the 11-dimensional
theory of the form (2.3) with
\be
ds^2_{10} = ds^2_4 + e^{2\phi/3} \sum^9_{i=4} dy^i \ dy^i 
\ee
where $y_i , \ i = 4, 5, \cdots , 9$ are flat directions which will be
compactified on a six-torus of volume $V_6$. Plugging (2.4) into the
action (2.1) gives
\be
S_{10} = {1 \over 2\pi g^2} {V_6 \over (2\pi)^6} \int d^4 x
\sqrt{-G_4} \left[ R_4 - {2 \over 3} (\nabla \phi)^2 - {1 \over 4}
e^{2\phi} F^2_4 \right]
\ee
Apart from some rescalings of the dilaton and the gauge potential, and
the different signature of the metric, this is exactly the action
whose solutions have been derived in \cite{thirtyseven, thirtyeight,
thirtynine}. These solutions have been further discussed in different
contexts in \cite{forty, fortya, fortyb}. So we may essentially read
off the desired solution from these works. For our purposes here we
will closely follow \cite{forty}.

The general spherically symmetric, asymptotically flat and
time-independent black hole solution obtained in this way involves
three arbitrary parameters, apart from the volume $V_6$ of the compact
6-dimensional torus. These may be taken to be the total Ramond-Ramond
``electric'' charge $Q$, the corresponding ``magnetic'' charge $P$ and
the ADM mass $M$ of the black hole. In terms of these parameters the
solution may be written as
\beq
ds^2_4 &=& - H_3 (H_1 H_2)^{-{1 \over 2}} dt^2 + (H_1 H_2)^{1 \over 2}
H^{-1}_3 dr^2 + (H_1 H_2)^{1 \over 2} d \Omega^2 , \\ [3mm]
A_\mu dx^\mu &=& - 2 \left[ {Q \over H_2} (r - {\lambda  \over 3}) dt
+ P (1 - \cos \theta) d \hat\phi \right] , \\ [3mm]
e^{4 \phi/3} &=& H_2 / H_1 ,
\eeq
where $H_1, H_2, H_3$ are the following functions of $r$ :
\beq
H_1 (r) &=& (r - \lambda/3)^2 - {2 \lambda P^2 \over \lambda - 3 G_N
M} , \\ [3mm]
H_2 (r) &=& (r + \lambda/3)^2 - {2 \lambda Q^2 \over \lambda + 3 G_N
M} , \\ [3mm]
H_3 (r) &=& (r - G_N M)^2 - (G^2_N M^2 - P^2 - Q^2 + \lambda^2/3) .
\eeq
In the above $G_N \equiv 8\pi^6 g^2/V_6$ is the 4-dimensional Newton's
constant and the parameter $\lambda$ is the ``dilaton charge'' defined
by the asymptotic behaviour
$$
\phi \rightarrow {\lambda \over r} + 0 ({1 \over r^2}) \ , \ \ \ \ \ r
\rightarrow \infty .
$$
The parameter $\lambda$ is not independent, but depends on $Q, P$ and
$M$ through the constraint
\be
{Q^2 \over \lambda + 3G_NM} + {P^2 \over \lambda - 3G_NM} = {2 \over
9} \lambda .
\ee
There exists a more convenient parametrization \cite{thirtyseven} which uses
three independent parameters, $q, p$ and $a$, in terms of which the
constraint (2.12) is automatically satisfied.  The physical parameters
$Q, P$ and $M$ are related to these by
\beq
G_NM &=& {1 \over 4} (q + p) , \ \ \lambda = {3 \over 4} (q - p) ,
\nonumber \\ [3mm]
Q^2 &=& {q \over 4} \left({q^2 - a^2 \over q + p}\right) , \ \ P^2 =
{p \over 4} \left({p^2 - a^2 \over q + p}\right) .
\eeq

The inner and outer horizons of the solution are defined by the zeroes
of $H_3 (r)$. These occur at $r = r_\pm$, given by
\beq
r_\pm &=& G_NM \pm (G_N^2 M^2 - P^2 -Q^2 + \lambda^2/3)^{1/2}
\nonumber \\ [3mm]
&=& {1 \over 4} (q + p) \pm {a \over 2} 
\eeq
We see that $a$ parametrizes devation from extremality. 

The entropy, $S$, and the Hawking temperature, $T_H$, of the black
hole are given by
\beq
S &=& {\pi \over G_N} \left(H_1 (r_+) H_2(r_+)\right)^{1/2} = {\pi
\over 2G_N} (pq)^{1/2} {(p+a)(q+a) \over (p+q)} \\ [3mm]
T_H&=& {1 \over 4\pi} (r_+ - r_-) (H_1 (r_+) H_2(r_+))^{-1/2} = {a \over 2\pi}
(pq)^{-1/2} \left({(p+a) (q+a) \over (p+q)}\right)^{-1} \nonumber \\
[2mm]
\eeq
The geometry, thermodynamics and other properties of this solution
have been extensively discussed in \cite{forty} to which we refer the
reader for details. Here we will only briefly discuss 
two special cases. 
\begin{enumerate}
\item[{(i)}] \underbar{Reissner-Nordstr\"om Solution} \\
~~~~~This requires a constant dilaton which is obtained for $\lambda =
0$. The constraint (2.12) (alternatively, the parametrization (2.13))
then requires $Q = P$, i.e. in this case the solution requires the
``electric'' and ``magnetic'' charges to be equal. The ADM mass of the
corresponding extremal solution is given by $G_NM = \sqrt{2} \ Q =
\sqrt{2} \ P$. This is the solution discussed in \cite{thirtytwo}. As we shall
see later, comparison with matrix theory requires the ratio of
electric to magnetic charges, $Q/P$, to be large. Such a comparison is
clearly not possible for the Reissner-Nordstr\"om case.
\item[{(ii)}] \underbar{The general extremal solution} \\
~~~~~This is obtained by setting $r_+ = r_-$. In the parametrization
(2.13) this implies that the parameter $a$ vanishes for the extremal
case, as is obvious from the second equality of (2.14). Using $a=0$ in
(2.13) one may solve for $q, p$ and $M$ in terms of the two
independent parameters $Q$ and $P$. For the ADM mass of the extremal
black hole one gets
\be
G_N M_{\rm ext} = {1 \over 2} (Q^{2/3} + P^{2/3})^{3/2} .
\ee
Similarly, from (2.15) one gets
\be
S_{\rm ext} = {2\pi \over G_N} QP
\ee
Now, in terms of the integer normalized D0-brane and D6-brane charges,
$Q_0$ and $Q_6$, corresponding respectively to $Q$ and $P$, we have
\cite{thirtytwo}
\be
Q = {Q_0 \over 4M_6} = 2 G_N M_0 Q_0 \ , \ \ \ P = {Q_6 \over 4M_0} =
2G_N M_6 Q_6 ,
\ee
where
\be
M_6 = {1 \over g} {V_6 \over (2\pi)^6} \ , \ \ \ M_0 = {1 \over g} ,
\ee
are respectively the mass of a single D6-brane and a single D0-brane.
Rewriting (2.17) and (2.18) in terms of $Q_0$ and $Q_6$, we get
\beq
M_{\rm ext} &=& \{(Q_0 M_0)^{2/3} + (Q_6 M_6)^{2/3}\}^{3/2} ,
\\ [3mm]
S_{\rm ext} &=& \pi \ Q_0 Q_6 .
\eeq
\end{enumerate}

It follows from (2.21) that the ADM mass of the extremal black hole is greater
than the sum of masses of $Q_0$ D0-branes and $Q_6$ D6-branes. So it
is unstable against decay into infinitely separated branes. As discussed in
\cite{fortyone}, even extreme black holes, which have a zero Hawking
temperature, can decay in theories in which there exist particles with
charges greater than their masses. This is so in the present case
since, in 4-dimensional Planck units, both types of branes have
charges which are twice their respective masses, as can be seen from
(2.19). Now, for large values of $Q_0 (Q_6)$, a WKB estimate for the
rate of decay \cite{fortyone} by the emission of D0-branes (D6-branes) is
$e^{-kQ_0} (e^{-kQ_6})$, where $k$ is a constant of order unity. Hence
black holes of this type with large values of the two charges are
long-lived states. 

A microscopic picture of this extremal black hole in terms of a bound
state of $Q_0$ D0- and $Q_6$ D6-branes has been discussed in
\cite{thirtytwo}. As argued there, the degrees of freedom responsible for the
entropy of the black hole are the fermionic modes in the Ramond sector
of the 0-6 strings. There modes are massless when the D0-brane is
sitting on top of the D6-brane. Using there modes a picture of the
bound state has been built and a counting of the
microscopic states of the bound system done in \cite{thirtytwo}.
Upto a constant of order unity, the logarithm of the degeneracy of
microstates is identical to the entropy of the black hole given by
(2.22). 

\section{MICROSCOPIC STRUCTURE OF THE GENERAL BLACK HOLE SOLUTION}

In this section we will present a new parametrization of the general
black hole solution in terms of four numbers $Q_0$, $\bar Q_0$, $Q_6$
and $\bar Q_6$ which we shall trade for the four parameters $Q$, $P$,
$M$ and $V_6$. As we shall see, in terms of these numbers the
expression for the entropy of the black hole takes the very simple
form
\be
S = \pi (Q_0 + \bar Q_0)(Q_6 + \bar Q_6)
\ee
which is valid arbitrarily far from extremality and is suggestive of an 
underlying microscopic brane and antibrane structure.

We begin by introducing the boost parameters $\alpha$ and $\beta$ as
follows:\footnote{The parametrization (2.13) of the general black hole
solution restricts the sum $(p + q)$ to be positive. This allows for
one of these two parameters to be negative. The following
parametrization corresponds to the case in which $q$ and $p$ are both
positive.}
\be
q = a \cosh\alpha,~~p = a \cosh\beta.
\ee
From (2.13) we then see that 
\be
Q = {a \over 2} (1 + {\cosh\beta \over \cosh\alpha})^{-1/2} \sinh\alpha
\ee
\be
P = {a \over 2} (1 + {\cosh\alpha \over \cosh\beta})^{-1/2} \sinh\beta
\ee
Now, analogous to (2.19) we wish to write (3.3) and (3.4) in the form 
\be
Q = 2 G_N M_0 (Q_0 - \bar Q_0),~~P = 2 G_N M_0 (Q_6 - \bar Q_6).
\ee
This suggests that we define the four numbers $Q_0$, $\bar Q_0$, $Q_6$
and $\bar Q_6$ as follows:
\be
Q_0 = {a \over 4 G_N M_0} (1 + {\cosh\beta \over \cosh\alpha})^{-1/2}
({e^\alpha + c \over 2})
\ee
\be
\bar Q_0 = {a \over 4 G_N M_0} (1 + {\cosh\beta \over \cosh\alpha})^{-1/2}
({e^{-\alpha} + c \over 2})
\ee
\be
Q_6 = {a \over 4 G_N M_6} (1 + {\cosh\alpha \over \cosh\beta})^{-1/2}
({e^\beta + d \over 2})
\ee
\be
\bar Q_6 = {a \over 4 G_N M_6} (1 + {\cosh\alpha \over \cosh\beta})^{-1/2}
({e^{-\beta} + d \over 2})
\ee
where at this stage $c$ and $d$ are arbitrary. These are partly fixed
by the requirement that the entropy of the extremal black hole,
(2.22), be correctly reproduced by the above parametrization. This
gives $c = d = 1$ in the extremal limit. We will make this simple
choice for $c$ and $d$ even away from extremality. Substituting (3.6)
- (3.9) for $c = d = 1$ in (2.15) then gives (3.1) for the entropy of
the general nonextremal black hole.

We can also express the mass $M$ of the black hole and the volume
$V_6$ of the internal $T^6$ in terms of the numbers $Q_0$, $\bar Q_0$,
$Q_6$ and $\bar Q_6$. Using (3.6) - (3.9) with $c = d = 1$, we get
\be
M = \left[\left(M_0\;{Q_0^2 + \bar Q_0^2 \over Q_0 + 
\bar Q_0}\right)^{2/3} +
\left(M_6\;{Q_6^2 + \bar Q_6^2 \over Q_6 + \bar Q_6}\right)^{2/3}\right]
^{3/2},
\ee
\be
{M_6 \over M_0} = {V_6 \over (2 \pi)^6} = \left[{Q_0 \bar Q_0 \over 
Q_6 \bar Q_6}\; {Q_6^2 + \bar Q_6^2 \over Q_0^2 + \bar Q_0^2}\right]
^{3/2} \left[{Q_0^2 + \bar Q_0^2 \over Q_0 + \bar Q_0}\right]\;
\left[{Q_6^2 + \bar Q_6^2 \over Q_6 + \bar Q_6}\right]^{-1}.
\ee
For completeness we also give below the expressions for the nonextremality
parameter, $a$, and the Hawking temperature, $T_H$:
\be
a = {1 \over M_6} \;{Q_0 \bar Q_0 \over Q_0 + \bar Q_0}
\left[1 + {Q_0 \bar Q_0 \over Q_6 \bar Q_6}\;
{Q_6^2 + \bar Q_6^2 \over Q_0^2 + \bar Q_0^2}\right]^{1/2},
\ee
\be
T_H = {2 M_0 \over \pi} {Q_0 \bar Q_0 \over (Q_0 + \bar Q_0)^2 
(Q_6 + \bar Q_6)}\left[1 + {Q_0 \bar Q_0 \over Q_6 \bar Q_6}\;
{Q_6^2 + \bar Q_6^2 \over Q_0^2 + \bar Q_0^2}\right]^{1/2}.
\ee
The extremal limit discussed in the previous section is obtained for
$\bar Q_0 \to 0$ and $\bar Q_6 \to 0$.\footnote{Other extremal limits 
can be obtained by
letting any one of the pairs of charges $(Q_0,Q_6)$, $(Q_0,\bar Q_6)$
and $(\bar Q_0,Q_6)$ vanish.}

The mass formula (3.10) generalizes (2.21) to the nonextremal case and
apparently corresponds to a collection of {\it interacting} branes and
antibranes. Unlike the extremal case, however, the mass of a
nonextremal black hole is not necessarily more than the sum of the
masses of its constituent branes and antibranes. In fact, for fixed
$Q$ and $P$, as we move away from extremality by adding branes and
antibranes, the mass of the black hole increases at a much slower rate
than the sum of masses of its constituents. This indicates that the
binding becomes tighter as more and more antibranes are added to the
system of branes that constitutes the extremal black hole. An extreme
example of this is the neutral Schwarzschild black hole. It can be
easily seen from (3.10) that the mass of Schwarzschild black hole is a
factor of $1/\sqrt{2}$ smaller than the sum of masses of its
constituents. The Schwarzschild black hole, therefore, appears to be a
truly bound state of its constituent branes and antibranes, unlike the
extremal black hole. It Hawking decays, however, because a brane and
the corresponding antibrane can annihilate when they overlap. This
decay can be neglected in the classical limit for large black
holes. In the microscopic picture this seems to indicate that for
large black holes the probability for branes (and antibranes) to
overlap is small, presumably because the average separation between
branes (and antibranes) grows with the size of the black hole.

To end this section we mention that the formulae (3.1), (3.10) and
(3.13) respectively for entropy, mass and the Hawking temperature of
the black hole are invariant under the conjectured U-duality group,
$E_{7(7)} (\bf Z)$, for the string theory under discussion
\cite{fortyonea}. This follows from the fact that the charges 
we are considering are inert under S-duality and are interchanged by
T-duality. Moreover, the latter also interchanges $M_0$ and
$M_6$. Using this and (3.11) one can then easily see the duality
invarince of the entropy, mass and Hawking temperature.

\section{D0-BRANE PROBE IN THE EXTREMAL BLACK HOLE BACKGROUND}

In this section we will consider the motion of a slowly moving
D0-brane probe in the presence of the extremal black hole from the
point of view of classical closed string theory. This is governed by
the action \cite{nine}
\be
S_{\rm Probe} = - {1 \over g} \int d\tau e^{-\phi} |{ds_{4,\rm ext} \over 
d\tau}| + {1 \over g} \int d\tau A_\mu {dx^\mu \over d\tau}
\ee
where $\tau$ parametrizes the trajectory of the probe D0-brane and $ds_{4, \rm
ext}$ refers to the line element in (2.6) specialized to the extremal case.
In the static gauge, the above action becomes
\be
S_{\rm Probe} = - {1 \over g} \int dt \left[ K_1^{1/2} K_2^{-1} (1 -
K_1 K_2 v^2)^{1/2} + 2 Q K_2^{-1} K_3 \right]
\ee   
In writing (4.2) we have introduced the variable $\rho = (r -
G_NM_{\rm ext})$ to bring the velocity $v$ of the probe into the
standard form $v^2 = {\dot \rho}^2 + \rho^2 {\dot \Omega}^2$. We have also 
dropped a term linear in velocity in the action since we will not be
considering this term when we compare with the matrix theory result.
The functions $K_1, K_2$ and $K_3$ are given by
\beq
K_1 (\rho) &=& 1 + {2 P \over \rho} \sqrt{f^2 + 1} +
{2 P^2 \over \rho^2} f^2, \\ [3mm]
K_2 (\rho) &=& 1 + {2 P \over \rho} f^2 \sqrt{f^2 + 1} +
{2 P^2 \over \rho^2} f^4, \\ [3mm]
K_3 (\rho) &=& {1 \over \rho} \left( 1 + {P \over \rho} \sqrt{f^2 +
1}\right), 
\eeq
where the parameter $f$ measures the ratio of the charges
\be
f^3 \equiv {Q \over P} = {Q_0 M_0 \over Q_6 M_6}. 
\ee

For small velocities we may expand the effective potential seen by a
D0-brane scattering off the black hole in powers of $v$. At large
distances $\rho$, this may be further expanded in powers of $\rho$. In
the present case, this latter expansion is identical to loop expansion
in the string coupling. Thus, the 1-loop effective potential, correct
to order $v^4$, is given by
\beq
V^{(1 {\rm -loop})}_{\rm eff} &=& - {Q_6 \over 2\rho} \bigg[ \{ (f^2 -
{1 \over 2}) \sqrt{f^2+1} - f^3 \} + {3v^2 \over 4} \sqrt{f^2+1}
\nonumber \\ [3mm]
&& + {v^4
\over 8} (f^2 + 5/2) \sqrt{f^2+1} + 0 (v^6)\bigg]
\eeq
Note that the black hole solution given in \cite{thirtytwo} corresponds to
$f=1$. The expression for $V_{\rm eff}^{(1 {\rm -loop})}$ given above
agrees with that given in \cite{thirtyfive} for $f=1$ to the approximation
considered there.

\section{THE YANG-MILLS SOLUTION}

It is known that both the short-range and long-range potentials
between a D0-brane and a D6-brane are repulsive. It is, therefore, not
possible to form a $0 + 6$ bound state without putting extra energy
into the system. In the Yang-Mills picture this is reflected in the
fact that energetically a D0-brane as an ``instanton'' in a D6-brane
gauge theory would prefer to shrink to a point and then move away from
the D6-brane. Nevertheless, as shown in \cite{thirtythree}, a Yang-Mills
configuration on $T^6$ corresponding to 4 D6-branes and 4
D0-branes exists which is classically stable at least to quadratic order.

We now look for a Yang-Mills configuration on $T^6$ for generic values
of the charges $Q_0$ and $Q_6$. As in \cite{thirtythree}, this may be
obtained by solving the following equations in a ${\cal U}(Q_6)$ gauge theory:
\be
F = \eta \ast (F \wedge F)
\ee
under the constraints
\be
\int_{\rm 2-cycle} Tr \ F = 0 \ , \ \ \ \int_{\rm 4-cycle} Tr (F
\wedge F) = 0
\ee
and
\be
Q_0 = {1 \over 6 (2\pi)^6} \int_{T^6} Tr (F \wedge F \wedge F)
\ee
Here $F$ is the 2-form field strength on $T^6$ and $\ast$ in (5.1)
denotes Hodge dual. Also, $\eta$ is a Lagrange multiplier which
forces the constraint (5.3). This constraint
ensures that the 
configuration has $Q_0$ D0-brane charge. The other constraints, (5.2),
ensure that the configuration has vanishing D2- and D4-brane charges.

As in \cite{thirtythree} we choose a solution with constant field strength with
only the following nonvanishing components in $T^6$:
\be
F_{45} = \gamma \nu_1 \ , \ \ F_{67} = \gamma \nu_2 \ , \ \ F_{89} =
\gamma \nu_3 
\ee
where $\gamma$ is an arbitrary constant and $\nu_1, \nu_2$ and $\nu_3$ are
diagonal $Q_6 \times Q_6$ matrices. According to (5.2) they must
satisfy
\be
Tr \nu_i = 0 \ , \ \ \ Tr (\nu_i \nu_j) = 0 \ (i \neq j) \ , \ \ i, j = 1,
2, 3
\ee
Moreover, the equation of motion (5.1) implies that the entries in all
the three $\nu_i$'s must be equal in magnitude, which we may set equal
to identify, without any loss of generality, because of the presence
of the parameter $\gamma$ in (5.4). The equations of motion are then
satisfied if
\be
\nu_i \nu_j = |\epsilon_{ijk}| \nu_k
\ee
A solution to (5.6) satisfying (5.5) has been presented in \cite{thirtythree}
for the smallest possible value of $Q_6$, which is 4. The simplest
generalization of this to arbitrary $Q_6$ is to just repeat the
$Q_6=4$ solution an arbitrary number of times. That is, a solution
with $Q_6 = 4n$ is given by
\beq
\nu_1 &=& {\rm diag} \left((1, 1, -1, -1), (1, 1, -1, -1), \cdots , n {\rm
times}\right) , \nonumber \\ [3mm]
\nu_2 &=& {\rm diag} \left((1, -1, -1, 1), (1, -1, -1, 1), \cdots , n {\rm
times}\right) , \nonumber \\ [3mm]
\nu_3 &=& {\rm diag} \left((1, -1, 1, -1), (1, -1, 1, -1), \cdots , n {\rm
times}\right) .
\eeq
Note that in the above the order in which the individual entries in
the diagonal matrices appear may be changed without affecting the
solution, provided identical changes are made in all $\nu_i$'s. Thus,
the following is an equivalent solution:
\beq
\nu_1 &=& {\rm diag} \left((1, 1, \cdots n {\rm times}), (1, 1, \cdots n {\rm
times}), (-1,-1, \cdots n {\rm times}) (-1, -1, \cdots n {\rm
times})\right) , \nonumber \\ [3mm] 
\nu_2 &=& {\rm diag} \left((1, 1, \cdots n {\rm times}), (-1, -1, \cdots n {\rm
times}), (-1,-1, \cdots n {\rm times}) (1, 1, \cdots n {\rm
times})\right) , \nonumber \\ [3mm] 
\nu_1 &=& {\rm diag} \left((1, 1, \cdots n {\rm times}), (-1, -1, \cdots n {\rm
times}), (1, 1, \cdots n {\rm times}) (-1, -1, \cdots n {\rm
times})\right). \nonumber \\ [2mm]  
\eeq
It is this form of the solution that will be convenient for the
scattering calculation done in the subsequent sections. 

Just as in the case of the configuration with $Q_6=4$ discussed in
\cite{thirtythree}, all supersymmetries are broken in the present configuration
as well. Since that is the case, one expects the energy of this
configuration to exceed the minimal BPS energy for the $0+6$ system.
In fact, using the Born-Infeld formula for diagonal field strengths,
we get for the energy of the above Yang-Mills configuration
\beq
E_{YM} &=& {1 \over g(2\pi)^6} \int_{T^6} Tr \sqrt{
\det(\delta_{\mu\nu} + F_{\mu\nu})} \nonumber \\ [3mm]
&=& M_6 Tr \{ (1 + F^2_{45}) (1 + F^2_{67}) (1 + F^2_{89})\}^{1/2}
\nonumber \\ [3mm]
&=& M_6 Q_6 (1 + \gamma^2)^{3/2}
\eeq
Now, from (4.3) we deduce that the parameter $\gamma$ is related to $Q_0$
and $Q_6$ by
\be
Q_0 M_0 = \gamma^3 Q_6 M_6 .
\ee
From this and (3.6) we see that $\gamma = f$. Moreover, using this in
(4.9) we get
\be
E_{YM} = \{(M_0 Q_0)^{2/3} + (M_6 Q_6)^{2/3}\}^{3/2} .
\ee
This is precisely the same as the expression for $M_{\rm ext}$ in
(2.21). This result is surprising since the configuration of branes
that we are considering does not preserve any supersymmetry and so one
might have expected the mass of the bound state to get renormalized in
the strong coupling region which is the region in which supergravity
is the effective low energy theory. A deeper appreciation of this
result could be useful in understanding aspects of strongly coupled 
dynamics that do not rely entirely on supersymmetry.

\section{MATRIX THEORY CALCULATION}

In this section we will calculate the 1-loop effective potential for a
D0-brane probe scattering off the configuration of D0- and D6-branes
constructed in the previous section. By now there exist many
calculations of 1-loop effective potential of D-brane probes
scattering off various configurations of D-branes within the matrix
theory framework. The set-up for a 1-loop calculation of D0-brane --
D6-brane scattering has been formulated in \cite{fortytwo} and the
calculation done for the special case in which the three magnetic
fluxes on the D6-branes are equal. From (5.8) we see that we need this
calculation for a more general case. We now proceed to do this
calculation. 

The only dynamical degrees of freedom in matrix theory are D0-branes
and their dynamics is governed by the quantum mechanical action
\cite{thirteen} obtained by dimensional reduction of 10-dimensional
superYang-Mills action to 1-dimension:
\beq
S &=& {1 \over 2g} \int dt \ Tr \bigg\{ (D_t X_i)^2 + {1 \over 2}
[X_i, X_j]^2 - (\bar D_t A)^2 + \theta^T D_t \theta \nonumber \\ [3mm]
&& ~~~~ + i \theta^T \gamma^i [X_i, \theta] + 2 \partial_t C^\dagger D_t C
- 2 [C^\dagger, B_i] [X_i, C] \bigg\} .
\eeq
Here $g$ is the string coupling, $X_i (i=1, 2, \cdots 9)$ are matrix
valued space components of $A_\mu$ (the 10-dimensional gauge
potential), $A_0$ is the time-component of $A_\mu , \theta$ is a real
16-component spinor ($\theta^T$ is transposed in spinor indices
only), $\gamma^i (i=1,2, \cdots 9)$ are nine real symmetric $16 \times 16$
Dirac matrices satisfying $(\gamma^i)^2=1$ and, finally, $C, C^\dagger$ are
the ghost fields. $B_i$ is the background value for $X_i$ and we have
chosen $B_0 = 0$. Also, 
\beq
D_tX_i &=& \partial_t X_i - i [A, X_i] , \ \ D_t \theta = \partial_t
\theta - i [A, \theta] , \nonumber \\ [3mm]
D_t C &=& \partial_t C - i [A, C] , \ \ \bar D_t A = \partial_t A + i
[B_i, X_i] . 
\eeq
Note that the dimensional reduction has been done after background
gauge-fixing and ghost terms have been added to the 10-dimensional action.

\underbar{Background Configuration}

The above quantum mechanical action was studied in \cite{fortythree} in
connection with the 11-dimensional supermembrane and in \cite{fortyfour} in
the context of Dp-branes in matrix theory formuation of M-theory.  In
this latter framework, all Dp-branes are made of D0-brane constituents 
and can be obtained as classical configurations of the action (6.1).
The configuration we desire, which is expected to correspond to the
extremal black hole solution of sec. 2, consists of D6-branes with the
D0-branes appearing on them as magnetic fluxes.  A multiple six-brane
configuration with magnetic fluxes in (45), (67) and (89) directions is 
given by the following
\[
B_{4,6,8} = \pmatrix{Q_{1,2,3} & \, & 0 \cr \, & \, & \,
\cr 0 & \, & 0}, \,\,\, B_{5,7,9} =
\pmatrix{P_{1,2,3} & \, & 0 \cr \, & \, & \, \cr 0 & \, & 0}
\]
where the entry in the right lower corner is a single element one.
This entry is for the probe D0-brane.  For a single six-brane
$[Q_a,P_a] = ic_a, a = 1,2,3$, while for multiple six-branes the
$Q_a$'s and $P_a$'s have a further structure:
\[
Q_a = \pmatrix{Q^1_a && \cr & Q^2_a & \cr && \ddots}, \,\,\, P_a =
\pmatrix{P^1_a & & \cr & P^2_a & \cr && \ddots}
\]
where $[Q^1_a, P^1_a] = ic^1_a$, etc.  The six-branes are wrapped on a 
$T^6$ with volume $V_6$ which is assumed to be large since we will be
neglecting the effect of winding modes.  For this configuration to
correspond to our extremal black hole we need the upper index on
$Q_a$'s and $P_a$'s to run from 1 to $Q_6$.   

Now, the configuration of magnetic fluxes in (5.8) implies that the
$Q_6 (= 4n)$ D6-branes can be organised into 4 sets, each consisting of
$n$ D6-branes.  Each D6-brane in the first set carries magnetic fluxes 
$(F_{45}, F_{67}, F_{89}) = (f,f,f)$.  D6-branes in the other three
sets carry the fluxes 
$(f,-f,-f), (-f,-f,f)$ and $(-f,f,-f)$.  Thus a more suitable notation 
for the $Q_a$'s and $P_a$'s is $Q^{l,\alpha}_a, P^{l,\alpha}_a$
where $l = 1,2,3,4$ and $\alpha = 1,2,\ldots n$ and
$[Q^{l,\alpha}_a, P^{m,\beta}_a] = i\delta^{\alpha\beta} \delta^{lm}
c^{l,\alpha}_a$.  For the desired configuration we need to take
$c^{l,\alpha}_a = c^a_l$ to be independent of $\alpha$.
Moreover, the four triplets of numbers $\{c^a_l\} \equiv
\left(c^1_l, c^2_l, c^3_l\right) \equiv \vec c_l$
crrespond to the four triplets of fluxes listed above and so we may
write 
\beq
\vec c_l &=& c \vec \epsilon_l, \nonumber \\
\vec \epsilon_1 &=& (1,1,1), \, \vec \epsilon_2 = (1,-1,-1), \, \vec
\epsilon_3 = (-1,-1,1), \, \vec \epsilon_4 = (-1,1,-1)
\eeq
As we shall see, agreement with supergravity calculation requires $c = 
f^{-1} \rightarrow 0$.

Let us now consider a D0-brane probe scattering off this background in 
directions transverse to the D6-branes.  This is represented by the
appearance of the additional backgrounds
\[
B_1 = \pmatrix{0 & \, &  0 \cr \, & \, & \, \cr 0 & \, & vt}, \,\,\, 
B_2 = \pmatrix{0 & \, & 0 \cr \, & \, & \, \cr 0 & \, & b}
, \,\,\, B_3 = 0.
\]
The only nonzero entry in the above matrices is in the lower right
corner.  Here $v$ is the velocity of the D0-brane, assumed to be along $x^1$,
and $b$ is the impact parameter.  

\underbar{Fluctuations}

In order to compute the effective potential for the scattering of the
D0-brane off the configuration of D0- and D6-branes represented by the 
background values $B_i$, we need to insert these background values in
the action (6.1) and integrate out the fluctuations around the
background.  There are basically two types of fluctuations.  The first 
type are nonzero square matrices which are fluctuations around
$Q^{l,\alpha}_a$ and $P^{l,\alpha}_a$.  These represent open strings
connecting the various branes in the background configuration.
However, these do not contribute to the 1-loop potential, so we will
not consider them here any further.  They do, however, contribute to
the potential at 2-loop \cite{fortyfive} and beyond.  The other type
are the ones with nonzero values in the last column or row of the matrices
$X_i,\theta$, etc.  These represent open strings connecting the probe
D0-brane and the branes in the background configuration.  They are
the relevant fluctuations for the present calculation.  We shall
paramterize these fluctuations as follows.  Writing $X_i = B_i +
\sqrt g Y_i$, we have
\[
Y_i = \pmatrix{0 & \, & \phi_i \cr \, & \, & \, \cr \phi^+_i & \, & 0}.
\]
Similarly, 
\beq
A = \sqrt g \pmatrix{0 & \, & \phi_A \cr \, & \, & \, \cr \phi^+_A & \, & 
0}, \, \theta =
\sqrt g \pmatrix{0 & \, & \chi_\theta \cr \, & \, & \, \cr \chi^+_\theta & 
\, & 0}, \, 
C = \sqrt g \pmatrix{0 & \, & \chi_c \cr \, & \, & \, \cr \tilde\chi^T_c &
 \, & 0}. \nonumber 
\eeq
Note that, in the notation used earlier, the fluctuations 
have the index structure $\phi^{l,\alpha}_{\ldots}$ and
$\chi^{l,\alpha}_{\ldots}$.  It is also useful to paramterize the
background as
\[
B_i = \pmatrix{D_i & \, & 0 \cr \, & \, & \, \cr 0 & \, & d_i}.
\]

\underbar{Action for fluctuations}

We may now expand the action (6.1) around the background $B_i$.  The
terms linear in fluctuations vanish because $B_i$ is a solution of
equations of motion.  For a 1-loop calculation of the effective
potential it is sufficient to retain only the quadratic terms in the
fluctuations. The action for fluctuations is then given by the sum of
the following four pieces:
\beq
S_Y & = & \sum_{l,\alpha} \int d\tau \Big[\phi_i^{l,\alpha^\dagger} 
(2H_\tau + 2H^{l,\alpha} + b^2) \phi_i^{l,\alpha} \nonumber \\
&& \displaystyle\hbox{~~~~~} + 2i c^1_l (\phi_4^{l,\alpha^\dagger}
\phi_5^{l,\alpha} 
- \phi_5^{l,\alpha^\dagger} \phi_4^{l,\alpha}) \nonumber \\
&&\displaystyle\hbox{~~~~~} + 2ic^2_l
(\phi_6^{l,\alpha^\dagger}\phi_7^{l,\alpha} - 
\phi_7^{l,\alpha^\dagger} \phi_6^{l,\alpha}) \nonumber \\
&&\displaystyle\hbox{~~~~~} + 2i c^3_l (\phi_8^{l,\alpha^\dagger}
\phi^{l,\alpha}_9 - 
\phi_9^{l,\alpha^\dagger} \phi_8^{l,\alpha})\Big] \\
S_A &=& \sum_{l,\alpha} \int d\tau \left[\phi_A^{l,\alpha^\dagger}(2H_\tau +
2H^{l,\alpha} + b^2) \phi_A^{l,\alpha} + 2iv_E (\phi_1^{l,\alpha^\dagger}
Q_A^{l,\alpha} - \phi_A^{l,\alpha^\dagger} \phi_1^{l,\alpha})\right] \nonumber
\\ [3mm]
S_\theta &=& i\sum_{l,\alpha} \int d\tau \left[\chi_\theta^{l,\alpha^\dagger}
(\partial_\tau - \gamma^i (D^{l,\alpha}_i - d_i^{l,\alpha})) \,
\chi_\theta^{l,\alpha}\right], \\
S_c &=& \sum_{l,\alpha} \int d\tau \left[\phi^{l,\alpha^\dagger}_c (2H_\tau +
2H^{l,\alpha} + b^2) \phi^{l,\alpha}_c + \tilde \phi^{l,\alpha^\dagger}_c
(2H_\tau + 2H^{l,\alpha} + b^2) \tilde
\phi_c^{l,\alpha}\right].\nonumber \\ [2mm]
\eeq
In writing the above, we have already made the Wick rotation $t
\rightarrow i\tau, A \rightarrow -iA$ and $v_t \rightarrow
v_E\tau, v_E = i v$, since calculations are more conveniently
done in Euclidean space.  Also,
\beq
H_\tau & = & {1\over 2} \left(-\partial^2_\tau + v_E^2\tau^2\right)
\nonumber \\
H^{l,\alpha} &=& {1\over 2} \sum^3_{a=1}
\left[\left(P^{l,\alpha}_a\right)^2 +
\left(Q^{l,\alpha}_a\right)^2\right].
\eeq

Integrating out the various fluctuations gives a product of
determinants.  The ghost determinants cancel against those coming from 
$\phi^{l,\alpha}_2$ and $\phi^{l,\alpha}_3$ sectors.  The remaining
determinants in the bosonic sector involve the following operators
(obtained after diagonalizing in the $(A,\phi_1)$ and $(\phi_4,\phi_5)$, 
$(\phi_6,\phi_7)$ and $(\phi_8,\phi_9)$ sectors):
\beq
O^{l,\alpha}_{\tau\pm} & \equiv & 2H_\tau + 2H^{l,\alpha} + b^2 \pm
2v_E \longrightarrow {1 \over \sqrt 2} (\phi_1^{l,\alpha} \pm 
i\phi_A^{l,\alpha}) \nonumber \\
O^{l\alpha}_{1\pm} & \equiv & 2H_\tau + 2H^{l,\alpha} + b^2 \pm 2c^2_l 
\longrightarrow {1\over \sqrt 2} (\phi^{l,\alpha}_4 \pm
i\phi_5^{l,\alpha}) \nonumber \\
O^{l,\alpha}_{2\pm} & \equiv & 2H_\tau + 2H^{l,\alpha} + b^2 \pm
2c^2_l \longrightarrow {1 \over \sqrt 2} (\phi^{l,\alpha}_6 \pm
i\phi_7^{l,\alpha}) \nonumber \\
O^{l,\alpha}_{3\pm} & \equiv & 2H_\tau + 2H^{l,\alpha} + b^2 \pm 2c^3_l
\longrightarrow {1 \over \sqrt 2} (\phi_8^{l,\alpha} \pm
i\phi_9^{l,\alpha}) \nonumber
\eeq
Shown against each of these operators are the corresponding
``diagonal'' field combinations.  Thus, the bosonic sector gives rise 
to the following product of determinants:
\be
\prod^4_{l = 1} \prod^n_{\alpha = 1} \left\{\det O^{l,\alpha}_{\tau +} 
\det O^{l,\alpha}_{\tau -} \prod^3_{a=1} (\det O^{l,\alpha}_{a+} \det
O^{l,\alpha}_{a-})\right\}^{-1}
\ee
Now, all the above operators are diagonal in the oscillator number
representation of the (1-dimensional) harmonic oscillator for
$H_\tau$ and (3-dimensional) harmonic oscillator for $H^{l,\alpha}$. In this
representation it is clear, because of the form of (6.3), that the
eigenvalues of $H^{l,\alpha}$ do not depend on $l, \alpha$. In fact,
the determinants in (6.8) depend only on $l$ and this dependence comes
only from the mixing terms in the various operators. To see what this
dependence on $l$ entails consider, for example, the product
$$
\prod^4_{l=1} \prod^n_{\alpha =1} (\det O^{l,\alpha}_{1+} \det
O^{l,\alpha}_{1-})^{-1}
$$
We have, from (5.3),
\beq
O^{l,\alpha}_{1\pm} &=& 2H_\tau + 2H^{l,\alpha} + b^2 \pm 2c \ \ {\rm
for} \ \ l = 1, 2 \nonumber \\ [3mm]
&=& 2H_\tau + 2H^{l,\alpha} + b^2 \mp 2c \ \ {\rm for} \ \ l = 3, 4
\nonumber 
\eeq
In other words, the two operators involving $+2c$ and $-2c$ are each
repeated 4 times in the above product over $l$, giving 
$$
\left\{\det (2H_\tau + 2H + b^2 + 2c) \det (2H_\tau + 2H + b^2 -
2c)\right\}^{-Q_6} ,
$$
where we have used $4n = Q_6$. We have also used the fact that the
eigenvalues of $H^{l, \alpha}$ do not depend on $l, \alpha$ and,
dropped these indices on $H$. In this way one can show that the
product of determinants in (6.9) is equal to
\beq
&&\bigg\{\det (2H_\tau + 2H + b^2 + 2v_E) \det (2H_\tau + 2H + b^2 -
2v_E) \nonumber \\ [3mm]
&& \times (\det (2H_\tau + 2H + b^2 + 2c) \det (2H_\tau + 2H + b^2 -
2c))^3 \bigg\}^{-Q_6}
\eeq

To evaluate the fermionic determinant coming from (6.6), we first
``square'' the operator involved using the fact that $\gamma^3$ does not
appear in the operator $i(\partial_\tau - \gamma^i (D^{l,\alpha}_i - d^{l,
\alpha}_i))$. This gives the operator $(2H_\tau + 2H^{l,\alpha} + b^2
+ v_E \gamma^1 + {\displaystyle\Sigma^3_{a=1}} c^a_l \sigma^a)$ where
$\sigma^1 = i \gamma^4 \gamma^5 , \ \sigma^2 = i \gamma^6 \gamma^7$
and $\sigma^3 = i \gamma^8 \gamma^9$. This 
operator involves mixing coming from the $\gamma$-matrix term $(v_E \gamma^1 +
\vec c_l \cdot \vec\sigma)$. It can be diagonalised and its
eigenvalues are
\beq
&& \pm c (\epsilon^1_l + \epsilon^2_l + \epsilon^3_l)\pm v_E, \ \pm
c(\epsilon^1_l + \epsilon^2_l - \epsilon^3_l) \pm v_E , \nonumber \\ [3mm]
&& \pm c (\epsilon^1_l
- \epsilon^2_l + \epsilon^3_l) \pm v_E , \ \pm c (-\epsilon^1_l + \epsilon^2_l
+ \epsilon^3_l) \pm v_E \nonumber
\eeq
where all four combinations of the $\pm$ signs are allowed in each of
these, making a total of sixteen eigenvalues. Substituting the values
of $\vec\epsilon_l$ from (6.3), we see that for each of the values of
$l$ these eigenvalues reduce to either $\pm 3c \pm v_E$ (each
combination of $\pm$ signs occuring once) or $\pm c \pm v_E$ (each
combination of $\pm$ signs occuring three times). Thus, the product
over $l$ of the determinants in the fermionic sector also simplifies
and we get for the contribution of this sector
\beq
&&\bigg\{ (\det (2H_\tau + 2H + b^2 + 3c + v_E) \det (2H_\tau + 2H + b^2 +
3c - v_E) \times \nonumber \\ [3mm]
&\times& \det(2H_\tau + 2H + b^2 - 3c + v_E) \det (2H_\tau + 2H + b^2 - 3c
- v_E))^{1/2} \times \nonumber \\ [3mm]
&\times& (\det (2H_\tau + 2H + b^2 + c + v_E) \det (2H_\tau + 2H + b^2
+ c - v_E) \times \nonumber \\ [3mm]
&\times& \det (2H_\tau + 2H + b^2 - c +v_E) \det (2H_\tau + 2H + b^2 -
c -v_E))^{3/2} \bigg\}^{Q_6}
\eeq
Note that for $v_E = 0 = c$, all the determinants in (6.9) and (6.10)
become equal and these two expressions cancel against each other. This
fact would seem to indicate that the present nonsupersymmetric
configuration of branes becomes supersymmetric in the infinite
momentum frame which corresponds to $c = f^{-1} = 0$. 

Apart from the power $Q_6$, the expressions appearing in (6.9) and
(6.10) are exactly the ones that have been evaluated earlier in
\cite{fortytwo}. The result in the present case for the 1-loop long range
effective potential is just $Q_6$ times that obtained in that work.
For small $c$ and $v$, the potential correct to order $v^4$ is, then,
\beq
V^{\rm 1-loop}_{\rm matrix} &=& - {Q_6 \over \rho} \left( - {3c \over
16} + {3v^2 \over 8c} + {v^4 \over 16c^3} + 0 (v^6)\right) \nonumber
\eeq
Comparing this with (4.7) we see that it agrees with the leading term
in the former in the limit $f = c^{-1} \rightarrow \infty$. We also
see that in this limit and at low velocities the $v^4$ term is the
leading term (since all higher powers of $v$ are accompanied by the
same factor $c^{-3}$; this is easily seen from (4.2) -- (4.5)). Thus
we see explicitly that the scattering is dominated by D0-brane --
D0-brane scattering, and hence approaches a supersymmetric
configuration, in this limit. In fact, assuming that this continues to
be the case beyond the 1-loop term, we can use the 2-loop calculation
of \cite{twenty} to predict the form of the matrix theory 2-loop potential
in the present case. According to the calculation of \cite{twenty} the
$v^4$ term in the D0-brane -- D0-brane scattering does not get
renormalized at the 2-loop level. For the present case of a D0-brane
scattering off the extremal black hole this implies that at the 2-loop
level, in the limit $f = c^{-1} \rightarrow \infty$, the largest power
of $f$, in the low-velocity expansion, can only come with a term of
order $v^6$ or higher power of $v$. Using (4.2) -- (4.5), we can obtain
the 2-loop effective supergravity potential:
\beq
V^{\rm 2-loop}_{\rm eff} &=& {g Q^2_6 \over 4\rho^2} \bigg[ \{f^6 -
f^3 (f^2 - {1 \over 2}) \sqrt{f^2 + 1} - {3f^2 + 1 \over 8}\} - {3v^2
\over 16} (3f^2 + 1) \nonumber \\ [3mm]
&& ~~ - {3v^4 \over 32} \{4f^4 + {5 \over 2} (3f^2 + 1)\} - {v^6
\over 64} \{4f^6 + 36f^4 \nonumber \\ [3mm]
&& ~~ + {35 \over 2} (3f^2+1)\} + 0 (v^8)\bigg]
\eeq
In the limit $f \rightarrow \infty$, the largest power, $f^6$, comes
with $v^6$ (and this is the highest 
power of $f$ that comes with all the higher order terms in $v$).
Therefore, the supergravity calculation agrees with the predicted
behaviour of the 2-loop matrix theory potential!

\section{CONCLUDING REMARKS}

We have discussed a 4-dimensional nonsupersymmetric black hole
solution to low energy type IIA string theory compactified on a
six-torus. The solution is parametrized by three parameters which are
the ADM mass and D0- and D6-brane charges, in addition to the volume
of the six-torus. We have presented a new parametrization of
this solution which trades these four parameters for four numbers. In
terms of these numbers the entropy of the general nonextremal black
hole takes on a very simple U-duality invariant form which suggests an
underlying brane and antibrane structure. It would be useful to
develop this picture further since it is likely to yield further
information about strong coupling dynamics that does not rely entirely
on supersymmetry.

The extremal solution has only the two charges as
independent parameters, its mass being determined in terms of these.
It is unstable against decay by emission of the branes, but for large
values of the charges it is long lived.  In this case there is a
microscopic picture as a bound state of D0- and D6-branes. We have
obtained a configuration of D0- and D6-branes, which has energy equal
to the ADM mass of the extremal black hole, as a classical solution in
a ${\cal U}(Q_6)$ Yang-Mills theory on $T^6$. We considered the
corresponding brane configuration in matrix theory and computed the
1-loop effective potential for the scattering of a D0-brane probe off
this brane configuration. We found that this agreed with the
corresponding supergravity calculation in the limit in which the ratio
of the D0- to D6-brane charges is large. In fact, assuming that this
continues to be true beyond 1-loop order, we used the 2-loop calculation of
\cite{twenty} to make a prediction for the present case. As we have
seen, this prediction agrees with the corresponding 2-loop
supergravity calculation. Thus, although the extremal black hole
solution is not protected by supersymmetry, it would seem that an
appropriate limit exists in which the underlying brane configuration
approaches a supersymmetric state in a definite and calculable way.

One of the defining features of a black hole is the event horizon. A
1-loop matrix theory computation sees only the Newtonian potential
$\sim 1/\rho$ and misses the subleading corrections which depend on
the horizon size. To see the horizon size, one needs to go beyond a
1-loop computation. A 2-loop computation was done for the case of the
5-dimensional black hole in \cite{fifteen} and it disagrees with the
supergravity prediction, but the status of this disagreement is not
clear. In the present case, however, there are already some positive
indications about the 2-loop term, as we have argued at the end of the
last section. For these reasons it is clearly of interest to do a
complete calculation of the 2-loop term \cite{fortyfive}.

\end{document}